# Controlling the switching field in nanomagnets by means of domain-engineered antiferromagnets

E. Folven[1,*], J. Linder[2], O.V. Gomonay[3], A. Scholl[4], A. Doran[4], A.T. Young[4], S.T. Retterer[5], V.K. Malik[6], T. Tybell[1], Y. Takamura[6], and J.K. Grepstad[1]

[1] Department of Electronics and Telecommunications, Norwegian University of Science and Technology, NO-7491 Trondheim, Norway

[2] Department of Physics, Norwegian University of Science and Technology, NO-7491 Trondheim, Norway

[3] Institute of Physics and Technology, National Technical University of Ukraine, 03056 Kyiv, Ukraine

[4] Advanced Light Source, Lawrence Berkeley National Laboratory, Berkeley, CA 94720, USA

[5] Center for Nanophase Materials Science, Oak Ridge National Laboratories, P.O. Box 2008 MS 6123, Oak Ridge, TN 37831, USA

[6] Department of Chemical Engineering and Materials Science, University of California-Davis, Davis, CA 95616, USA

**Abstract:**

Using soft x-ray spectromicroscopy, we investigate the magnetic domain structure in embedded nanomagnets defined in $La_{0.7}Sr_{0.3}MnO_3$ thin films and $LaFeO_3/La_{0.7}Sr_{0.3}MnO_3$ bilayers. We find that shape-controlled antiferromagnetic domain states give rise to a significant reduction of the switching field of the rectangular nanomagnets. This is discussed in the framework of competition between an intrinsic *spin-flop* coupling and shape anisotropy. The data demonstrates that shape effects in antiferromagnets may be used to control the magnetic properties in nanomagnets.

**Main text:**

The coupling of an antiferromagnet to an adjacent ferromagnet may induce a unidirectional anisotropy, known as *exchange bias* [1], which is commonly exploited in spintronic devices [2,3]. This effect is utilized to achieve independent control of the magnetization in the different layers of magnetic tunnel junctions and spin valves used, e.g., in hard drive read heads and magnetic random access memory. More recently, the discovery of electric control of *exchange bias* has gained considerable attention [4,5]. This finding adds a degree of freedom to spintronic engineering. Furthermore, the increased coercivity typically arising from antiferromagnetic/ferromagnetic (AF/FM) coupling may help overcome superparamagnetism [6] in nanomagnets, thus allowing for information densities beyond present limitations to magnetic information storage.

While shape effects in ferromagnets are well understood and widely used to tailor their magnetic anisotropy, this is not the case for antiferromagnets. Limited control of the AF ground state restricts the possibilities for magnetic engineering in AF/FM bilayer systems. Previously, extrinsic properties such as interface roughness [7] have been invoked to tune the magnetic coupling. We have recently shown how the AF domain structure and interface spin alignment in AF/FM bilayer systems can be controlled using nanoscale patterning [8,9]. Furthermore, theoretical work has shown that the shape of an AF particle may introduce additional magnetic anisotropy via magnetoelastic coupling [10]. The magnetic easy axis is influenced by the presence of the surface, and as a result, the spontaneous strain associated with the orientation of the AF Néel vector at the surface is in general incompatible with that in the bulk. The internal stresses may relax either by formation of a domain structure or by reorientation of the AF Néel vector. In the simplest case of a rectangular nanomagnet with the edges parallel to two mutually orthogonal magnetic easy axes, the surface strain imposes a preferred direction of the AF Néel vector parallel to the long edge. This mechanism provides an additional tool for control of the AF ground state.

In this letter, we show that the competition between intrinsic *spin-flop* coupling in an AF/FM system and shape effects in the AF layer results in a reduced switching field in nanomagnets. We explain this finding in terms of stabilization of an additional uniaxial anisotropy in the FM layer. This demonstrates how the use of antiferromagnets with a tailored domain state opens up the possibility for a new approach to tune the magnetic anisotropy in nanomagnets.

We rely on a model system of thin films of 100 unit cells (u.c.) FM $La_{0.7}Sr_{0.3}MnO_3$ (LSMO) and AF/FM bilayers of 10 u.c. $LaFeO_3$ (LFO)/90 u.c. LSMO were grown epitaxially by pulsed laser deposition (PLD) on Nb-doped (0.05 wt %) (001)-oriented $SrTiO_3$ (STO) substrates using growth conditions reported previously [11]. The PLD growth was monitored *in situ* with reflection high-energy electron diffraction. Unit-cell intensity oscillations of the specular reflection were observed throughout the growth, and x-ray diffraction measurements showed that the thin films were fully strained to the in-plane lattice parameter of the substrate ($a$ = 3.905 Å) with LSMO and LFO out-of-plane lattice parameters of $(d_{001})_{pc}$ = 3.86 Å and $(d_{001})_{pc}$ = 4.03 Å (pseudocubic notation), respectively. Rocking curve widths for the $(001)_{pc}$ reflection were comparable to that of the substrate (FWHM < 0.02°). The film surface roughness was examined with atomic force microscopy, which showed step-and-terrace surfaces with sub-monolayer roughness on individual terraces. Rectangular nanomagnets (500 nm × 2 μm) with their edges oriented along in-plane $\langle 100 \rangle_{pc}$ directions were defined using $Ar^+$ ion implantation through a Cr hard-mask defined by electron beam lithography. The ion implantation serves to disrupt the structural and magnetic order in the AF/FM bilayer outside the regions shielded by the Cr hard-mask, leaving nanomagnets embedded in a paramagnetic matrix (for details see Refs. [12,13]).

Magnetic domain images of the AF/FM nanomagnets were obtained from x-ray magnetic linear/circular dichroism (XMLD/XMCD) measurements in combination with photoemission electron microscopy (PEEM), using the PEEM-3 microscope at the Advanced Light Source. The FM domain images were obtained by dividing PEEM images recorded with right-/left-handed helicity of the incident x-rays at the photon energy corresponding to the maximum XMCD signal, i.e., near the Mn $L_3$ absorption edge of LSMO. AF domain images were obtained by dividing PEEM images obtained using linearly polarized x-rays (s-polarization) at two different photon energies, corresponding to the two maxima of the Fe $L_2$-edge multiplet. A sample holder with an integrated electromagnet was used to impose small magnetic field pulses up to $H_{ext}$~190 Oe. The applied magnetic field was aligned parallel to the incident x-rays, with the samples mounted so that the x-ray incidence was parallel to either the [110] or [100] substrate directions. The measurements were performed at 110 K, i.e., well below $T_C$ ~ 270 K for LSMO in these samples. The magnetic yoke was saturated in a fixed direction prior to cooling, so that the remanent field set the magnetization of all 'bits' in one direction upon cooling the sample through $T_C$.

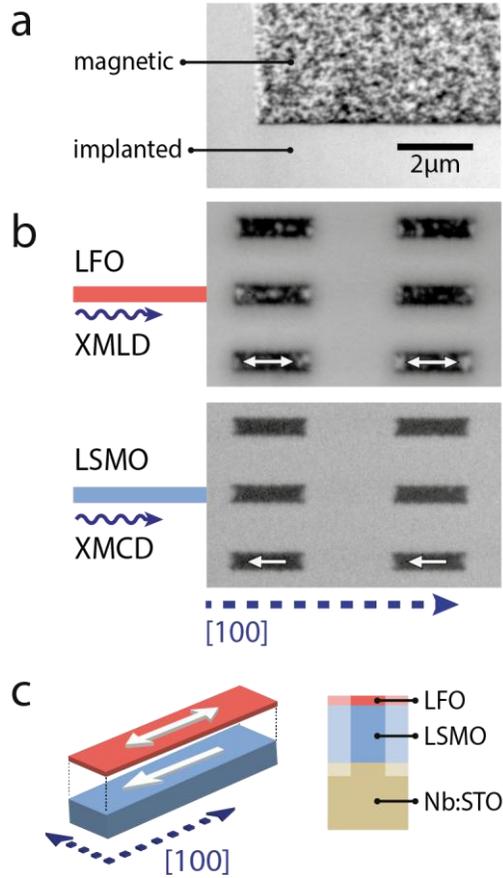

FIG. 1. (color online) (a) XMLD-PEEM image of the boundary between a patterned and un-patterned region in a single layer AF LFO thin film. We note the predominance of domains parallel to the patterned edge close to the boundary. (b) XMLD-PEEM (top) and XMCD-PEEM (bottom) images showing nanomagnets in the AF LFO and the FM LSMO layer, respectively. Edge-stabilized domains dominate the AF domain pattern and the magnetization in the FM is governed by shape anisotropy. (c) Schematic of the collinear spin structure in the bilayer nanomagnets and the cross-section of the embedded nanomagnets.

The XMLD-PEEM image in Fig. 1(a) shows the boundary between an extended patterned and un-patterned region in an AF LFO thin film. The region shielded during Ar+ ion implantation retains its AF domain structure while no magnetic signal is found within the implanted area. We note a predominance of dark domain contrast at the horizontal boundary of the AF region and correspondingly, a bright contrast at the vertical boundary of this region. This observation is understood in terms of edge-induced domain stabilization in the antiferromagnet. We have previously shown that edge-stabilized domains in nanostructures of widths less than 500 nm dominate the AF domain structure [8].

The ideal (001) surface of a G-type antiferromagnet such as LFO is magnetically fully compensated. Therefore, one would expect perpendicular alignment of the spins between the AF and FM layers [14] in coupled AF/FM heterostructures. Indeed, this *spin-flop* coupling was experimentally reported for extended LFO/LSMO thin films [15] and micrometer sized magnets

[16]. However, we have recently demonstrated that shape-induced domain stabilization may override this interface exchange coupling and force a collinear spin alignment in embedded LFO/LSMO nanomagnets below a certain critical width of approximately 500 nm [9]. In the rectangular nanomagnets (500nm×2µm) investigated in this study, edge-stabilized domains prevail (Fig. 1(b)). Thus, the AF Néel vector lies primarily parallel to the long edge in these rectangular 'bits'. This alignment implies parallel AF and FM spins in the bilayer nanomagnets, as the FM magnetization is also governed by shape and stabilized parallel to the long edge of the 'bits' which is evident from the homogenous dark domain contrast in the XMCD image in Fig. 1(b). Close inspection of the shape of the small grey patches barely visible at the short edges suggests an S-type domain structure [17], as expected for a shape anisotropy driven system. The mottled AF domain pattern seen in some 'bits' in Fig. 1(b) is interpreted as regions of the LFO layer where the *spin-flop* coupling persists or the AF spin axis is pinned by local defects. A schematic of the spin structure in the AF/FM nanomagnets investigated is shown in Fig. 1(c).

The switching characteristics of LSMO single-layer and LFO/LSMO bilayer nanomagnets were investigated by imposing 0-190 Oe magnetic field pulses *in situ* in the PEEM microscope followed by XMCD-PEEM imaging. The ~1 s field pulses were imposed immediately before image acquisition, so that the domain state was recorded in remanent conditions with zero applied field [18].

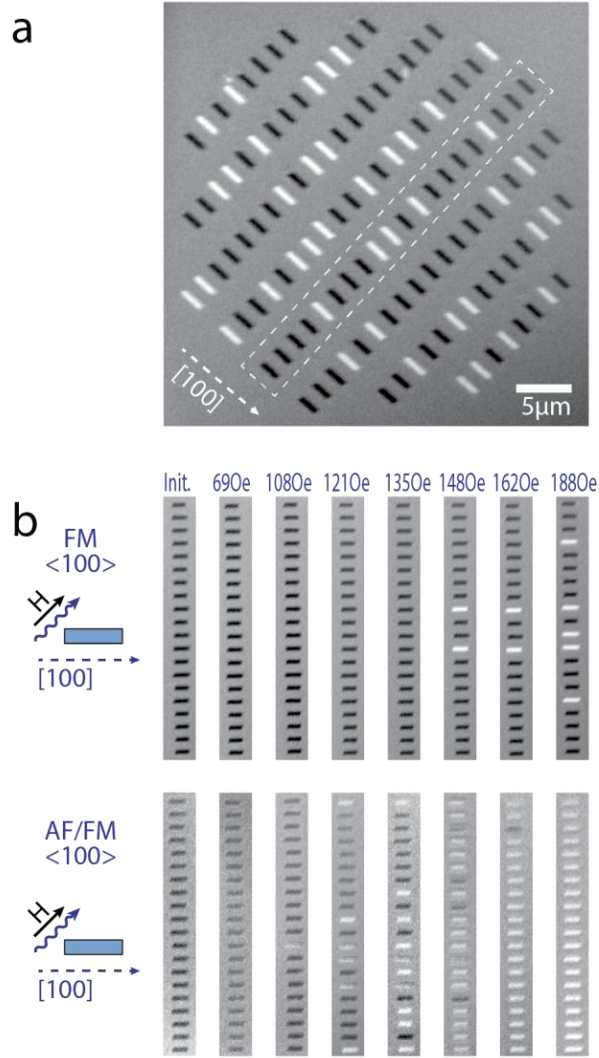

FIG. 2. (color online) (a) XMCD-PEEM images of an array of more than 100 nominally identical nanomagnets defined in FM LSMO; (b) selected column of this array after initial uniform magnetization and subsequent applied magnetic field pulses of increasing magnitude in the opposite direction, for LSMO single-layer (top panel) and LFO/LSMO bilayer nanomagnets (bottom panel). The experimental geometry appears from the schematics on the left.

Figure 2(a) shows the XMCD-PEEM image of a full array of nanomagnets defined in the LSMO single-layer. Figure 2(b) depicts the magnetization switching for nanomagnets defined in the LSMO single-layer (top panel), and corresponding data for nanomagnets defined in the LFO/LSMO bilayer (bottom panel). For each value of the applied magnetic field, Fig. 2(b) displays only one column of the full array of nanomagnets, as denoted with a dashed line in Fig. 2(a). The somewhat diffuse XMCD contrast in the PEEM images of the bilayer nanomagnets, as compared to those of the LSMO single-layer, is due to attenuation of the XMCD-PEEM signal by the thin (10 u.c.) LFO top layer. Even at the maximum applied field available in the PEEM microscope ($H_{max}$ = 190 Oe), only a small share of the nanomagnets defined in the LSMO single-layer switch, whereas nearly all of the bilayer nanomagnets undergo magnetization reversal, i.e., switch from black to white in Fig. 2(b). Thus, the presence of a thin AF layer gives rise to a substantial reduction in switching field for these bilayer magnets, a surprising observation as coupling to an antiferromagnet usually increases the switching field due to the additional drag [19]. Figure 3 plots the percentage of switched nanomagnets in the measured arrays (cf. Fig. 2(a)) versus applied field, comparing data for magnets defined in the LSMO single-layer and the LFO/LSMO bilayer. Data for the field applied along the [110] and [100] directions are plotted as closed and open symbols, respectively. A similar trend is observed between both field orientations. The switching field for AF/FM bilayer nanomagnets is reduced by approximately 30% compared to that for single layer nanomagnets. Within experimental error, the switching characteristics recorded for magnetic field pulses applied in the opposite direction did not show any signature of *exchange bias* for this system (see Supplementary, Fig S1 [20]).

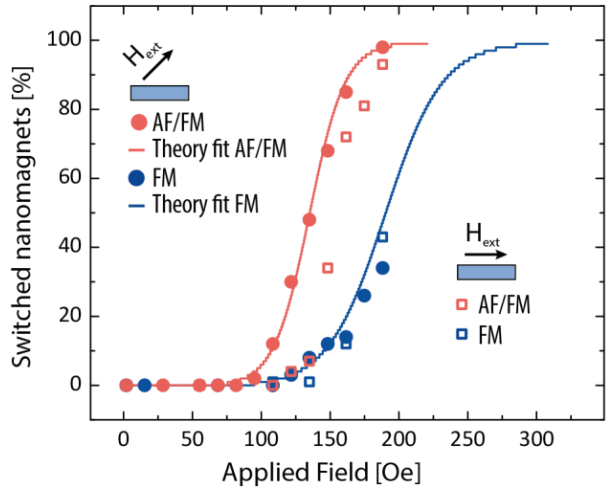

FIG. 3. (color online) Percentage of switched nanomagnets in the measured ensemble as a function of the applied magnetic field pulse for nanomagnets defined in the LSMO single-layer and the LFO/LSMO bilayer, respectively. The solid lines show numerical fits using the adopted model for magnetic switching of one hundred nanomagnets. We note that although each magnet switches abruptly, there is a statistical spread in the switching field for the ensemble.

The reduction in switching field for the bilayer nanomagnets is attributed to the shape-induced anisotropies in the AF and FM layers, which impose a ground state with collinear spin alignment and competes with the interface exchange coupling favoring perpendicular spin alignment. With the moderate fields required to switch the FM magnetization, we assume that the AF spins in the LFO layer remain aligned with the long edge of the rectangular nanomagnets under the applied field pulses. Thus, the interface exchange coupling will effectively act to reduce the energy associated with perpendicular orientation of the FM moments. The interface spin coupling thus adds a uniaxial contribution to the effective magnetic anisotropy of the FM layer perpendicular to the long edge of the rectangular nanomagnets, as depicted in Fig. 4(a,b).

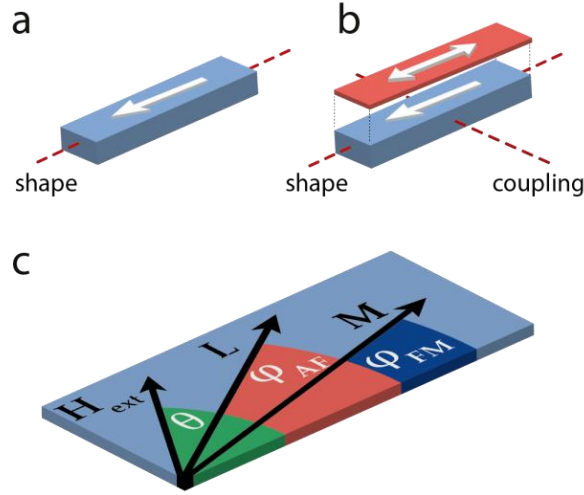

FIG. 4. (color online) Schematic showing the effective anisotropy axes imposed on the FM layer; (a) shape anisotropy for the LSMO single-layer, and (b) shape and interface coupling for the LFO/LSMO bilayer. (c) illustrates the angles defining the orientations of the external field, $H_{ext}$, the AF Néel vector, $L$, and the FM magnetization, $M$.

We find that the experimental data can be accounted for within a simple Stoner-Wohlfarth model, with the free energy $\mathcal{F}$ of the system written as:

$$\mathcal{F} = \mathcal{F}_{FM} + \mathcal{F}_{AF} + \mathcal{F}_{coupling}$$

where the three terms describe the free energy of the ferromagnet, the antiferromagnet, and the interlayer coupling, respectively. We have,

$$\mathcal{F}_{FM} = -\frac{1}{2} M_0 V H^{FM}_{shape} \cos^2 \phi_{FM} - M_0 V H_{ext} \cos(\phi_{FM} - \theta)$$

$$\mathcal{F}_{coupling} = \frac{1}{2} M_0 V H_{coupling} \cos^2(\phi_{FM} - \phi_{AF})$$

where $M_0$ is the saturation magnetization, $V$ is the volume of the FM region, and $H^{FM}_{shape}$, $H_{ext}$, and $H_{coupling}$ are the shape anisotropy field, external field, and the interface exchange coupling field, respectively. The angles $\phi_{FM}$ and $\phi_{AF}$ denote the orientation of the FM and AF order parameters $M$ and $L$ relative to the nanomagnet long axis, whereas $\theta$ is the angle between the external field and this axis (cf. Fig. 4(c)). While the interface exchange coupling $\mathcal{F}_{coupling}$ favors a perpendicular alignment of the spins in the LFO and LSMO layers, as observed experimentally for blanket films and larger micromagnets [9], shape anisotropy predominates and gives rise to a collinear alignment of **M** and **L** for the magnets displayed in Fig. 2. In these nanomagnets, the orientation of the FM moments is dictated by shape anisotropy. To a first approximation, the magnetocrystalline anisotropy along in-plane <110> easy axes for LSMO films under tensile strain [21] is therefore ignored in the present analysis.

To understand the reduced switching field for the AF/FM bilayer nanomagnets, we first note that the AF layer is nearly monodomain, so that the free energy may be written on the form [10],

$$\mathcal{F}_{AF} = -\frac{1}{2} M_0 V H_{shape}^{AF} \cos^2(\phi_{AF})$$

where $H_{shape}^{AF}$ is the shape anisotropy field for the AF layer. This relation implies that the AF order parameter **L** is aligned with the shape anisotropy easy axis for $\phi_{AF} = 0$. Inserting $\phi_{AF} = 0$ in the interface exchange coupling term, we see that the net effect of $\mathcal{F}_{coupling}$ is a reduced effective anisotropy,

$$H_{eff} = H_{shape}^{FM} - H_{coupling}$$

where the free energy to be minimized as a function of $\phi_{FM}$ becomes:

$$\mathcal{F}_{eff} = -\frac{1}{2} M_0 V H_{eff} \cos^2(\phi_{FM}) - M_0 V H_{ext} \cos(\phi_{FM} - \theta)$$

The interface exchange coupling to the AF layer reduces the required switching field required by a factor

$$R = \frac{H_{shape}^{FM}}{H_{shape}^{FM} - H_{coupling}} = \frac{1}{1 - H_{coupling}/H_{shape}^{FM}}$$

Comparison with the experimental data in Fig. 4 gives $R \simeq 1.5$, which leads to an estimate for $H_{coupling} \simeq \frac{1}{3} H_{shape}^{FM}$.

As the energy barrier $\Delta E$ between the two minima in the free energy of the ferromagnet is approximately four orders of magnitude larger than the thermal energy, $k_B T$, magnetization reversal should only happen if the applied field is large enough that $\Delta E$ changes sign, giving a step function profile for the magnetization direction vs. applied field (for details see Supplementary [20]). The gradual slope in the experimental data in Fig. 3 is attributed to the fact that we are dealing with an ensemble of nanomagnets with a certain variation in edge roughness, actual size, and density of defects, leading to variations in the required switching field. The individual nanomagnet will, however, switch abruptly. Numerical fits to the experimental data were obtained using the adopted model with the magnetic field applied along the [110] direction, shown as solid lines in Fig. 3, assuming a spread in switching field of ~35% for the FM single-layer and ~20% for the AF/FM bilayer. The reduced variation in switching field for the AF/FM bilayer nanomagnets indicates that the AF layer leads to more uniform switching of the nanomagnets. It should be noted, that a simple Stoner-Wohlfarth model, where the inherent anisotropy in the FM layer is presumed to be of uniaxial nature (i.e., predominated by shape, neglecting contributions from the biaxial magnetocrystalline anisotropy along the in-plane <110> axes), does not adequately reproduce the variations in switching behavior as a function of the angle $\theta$ of the applied field. In a system with mainly uniaxial anisotropy, the Stoner-Wohlfarth model predicts a large variation in switching field with field direction. This variation is reduced when higher order anisotropy terms are present [22-24]. The model including the biaxial magnetocrystalline anisotropy of LSMO is presented in Supplementary Information [20].

In conclusion, this work shows that substantial reduction in the switching field of nanomagnets can be obtained by engineering of the AF domain state through shape effects. We explain the result in terms of a competition between shape-induced anisotropy in the antiferromagnet and intrinsic *spin-flop* coupling across the AF/FM interface. This approach offers a new way to tailor the magnetic properties of AF/FM bilayer systems and should stimulate further research on a variety of topics where ultralow energy switching of nanomagnets is key.

**Acknowledgements**
Part of this work was carried out at the Center for Nanophase Materials Sciences, which is sponsored at Oak Ridge National Laboratory by the Office of Basic Energy Sciences, U.S. Department of Energy (DOE). The Advanced Light Source is supported by the Director, Office of Science, Office of Basic Energy Sciences, of the U.S. DOE under Contract No. DE-AC02-05CH11231. Partial funding for this experiment was obtained from the Research Council of Norway under Grants No. 205591, 216700, and 190086/S10 and from the National Science Foundation (DMR 0747896 and DMR 1411250).

Supplementary Information:

# Controlling the switching field in nanomagnets by means of domain-engineered antiferromagnets

E. Folven, J. Linder, O.V. Gomonay, A. Scholl, A. Doran, A.T. Young, S.T. Retterer, V.K. Malik, T. Tybell, Y. Takamura, and J.K. Grepstad

## Switching of AF/FM bilayer nanomagnets with positive and negative fields

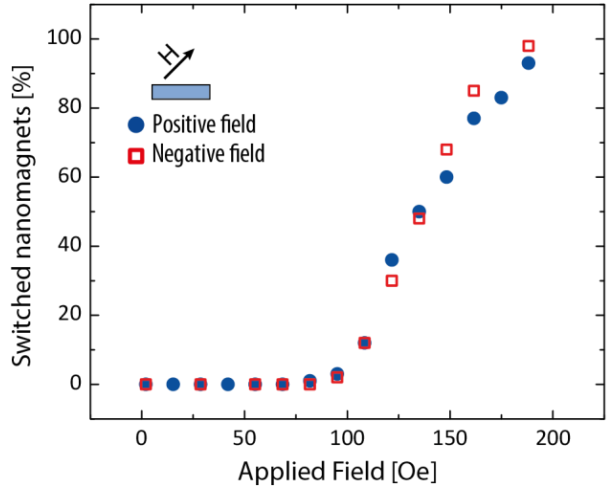

**Figure s5.** Percentage of switched nanomagnets as a function of the absolute value of the applied magnetic field pulse for nanomagnets defined in the LFO/LSMO bilayer.

## Considerations on thermally induced switching in nanomagnets

For the switching to take place on a reasonable timescale, the energy barrier $\Delta E$ between the two minima in the free energy of the ferromagnet must be comparable in magnitude to thermal energy $k_B T$. From the expression for the free energy, an analytical expression can be derived for the magnetization orientations $\phi_{FM,1}$ and $\phi_{FM,2}$ that represent the free energy minima:

$$\phi_{min,1(2)} = \pi/4 \mp \arccos\left(h \pm \sqrt{h^2 + 0.5}\right), \; h \equiv H/(2H_{eff}).$$

Similarly, magnetization orientations which maximize the free energy are $\phi_{max,1(2)} = \pi/4 \pm \arccos(h \pm \sqrt{h^2 + 0.5})$. From Fig. 3, switching of the nanomagnets in the FM single-layer commences for an applied pulse of $H \simeq 130$ Oe. From this value, we can estimate the magnitude of the energy barrier $\Delta E = \mathcal{F}(\phi_{FM} = \phi_{max}) - \mathcal{F}(\phi_{FM} = \phi_{min})$ that must be surmounted for the nanomagnets to switch between the minima. With nanomagnet volume $V = 500 \times 2000 \times 40\ nm^3$ and a saturation magnetization[2] $M_0 \simeq 4 \times 10^5$ A/m, we obtain an energy barrier of order $\Delta E \sim 10^{-17}$ J, as compared to thermal energy at $T = 100$ K, $k_B T \sim 10^{-21}$ J. Thus, the probability of thermally excited magnetization reversals, proportional to $e^{-\Delta E/k_B T}$, is vanishingly small and does not support reversal on any realistic timescale, even when taking into account the usual tunneling attempt frequency[3,4] on the order of $f_0 \sim 10^9$ Hz. Thus, magnetization switching should only happen if the applied field is large enough that $\Delta E$ changes sign, giving a step function profile for the magnetization direction vs. applied field.

To calculate the number of switched nanomagnets as a function of the applied field, we consider a model with $N_1$ nanomagnets being in the initial magnetization state and $N_2$ nanomagnets being in the state with opposite magnetization, i.e. the switched state. With $N = N_1 + N_2$ thus being the total number of nanomagnets, we have $N_1 (H=0)/N = n_1 (H=0) = 1$ as the initial condition, since all nanomagnets are magnetized in the same direction at zero field. The change in the fraction of nanomagnets $n_1$ as the field $H$ is tuned is then given by,

$$\frac{dn_1}{dH} = A[p_{2\to 1} - (p_{1\to 2} + p_{2\to 1})n_1]$$

using that $n_1 + n_2 = 1$. The tunneling probabilities are given by:

$$p_{1\to 2} = \exp\left(-\frac{\Delta E_1}{kT}\right), \quad p_{2\to 1} = \exp\left(-\frac{\Delta E_2}{kT}\right)$$

where $\Delta E_1$ and $\Delta E_2$ are the energy barriers separating transitions from state 1 to state 2 and vice versa. These barriers depend on the strength of the external field, making the above relations a differential equation that can only be solved numerically. The proportionality constant $A$ above is treated as a fitting parameter to the experimental data, while $k$ and are Boltzmann's constant and the temperature, respectively.

**Magnetization reversal in nanomagnets with higher order anisotropy terms**

In the absence of magnetocrystalline anisotropy, Stoner-Wohlfarth theory predicts that the required switching field is twice as large when applied along the magnetization axis compared to the field required when applied at an angle of 45 degrees. In the present measurements, we note that the switching field is indeed higher when the field is applied in the [100] direction, but far less than twice the magnitude of switching field when applied in the [110] direction.

In order to explain this deviation, we have included a biaxial magnetocrystalline anisotropy (along the in-plane <110> axes) in the free energy $\mathcal{F}$ by adding the term:

$$\mathcal{F}_{crystal} = -\frac{1}{2}M_0 V H_{crystal} \sin^2(2\phi_{FM})$$

In this case, an analytical procedure to calculate the ground-state magnetization angle is no longer available. However, the switching field can still be obtained by numerical computation. Since the magnetocrystalline anisotropy is expected to be weak compared to the shape anisotropy, we assume $H_{crystal} \ll H_{shape}$, so that the equilibrium magnetization remains near parallel to the shape anisotropy axis. As seen from Fig. s2, the switching field is twice as large for a [100]-oriented field compared to a [110]-oriented field when $H_{crystal} = 0$. However, upon increasing $H_{crystal}$, we note that the switching field for the two geometries, i.e., $\theta = \frac{3\pi}{4}$ and $\theta = \pi$, where $\theta$ denotes the orientation of the applied magnetic field relative to the long axis of the nanomagnet, approach one another. This is consistent with the experimental data for our samples, where the switching field for a [100]-oriented applied field is only slightly higher than for a [110]-oriented field.

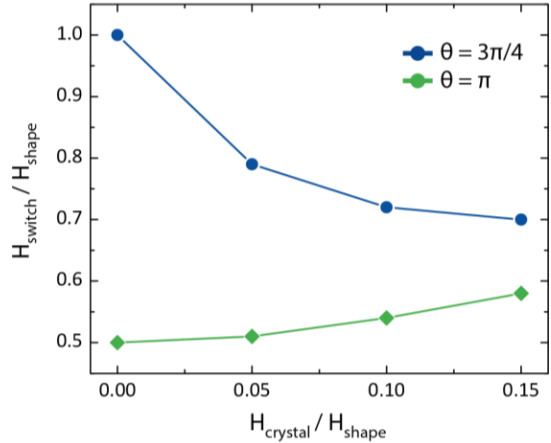

**Figure s2.** The normalized switching field as a function of magnetocrystalline anisotropy, (normalized to the shape anisotropy).